\documentstyle[12pt,psfig]{article}
\def\beq{\begin{equation}}
\def\eeq{\end{equation}}
\def\bce{\begin{center}}
\def\ece{\end{center}}
\def\bea{\begin{eqnarray}}
\def\eea{\end{eqnarray}}
\def\ben{\begin{enumerate}}
\def\een{\end{enumerate}}

\def\ni{\noindent}

\def\ms{\medskip}

\def\brr{\begin{array}}
\def\err{\end{array}}

\def\Sp{\mbox{Sp}\, }
\def\arg{\mbox{arg}\,}
\def\Re{\mbox{Re}\, }

\begin{document}
\baselineskip=17pt
{\large
\title{\bf Dynamical symmetry restoration for a higher-derivative 
four-fermion model in an external electromagnetic field}
\author{E. Elizalde$^{1,2,}$\thanks{E-mail: eli@zeta.ecm.ub.es,
elizalde@io.ieec.fcr.es}, \ 
S. P. Gavrilov$^{3,4,}$\thanks{E-mail: gavrilov@sergipe.ufs.br }, \ 
S. D. Odintsov$^{1,4,}$\thanks{E-mail: odintsov@tspi.tomsk.su}, \    
Yu. I. Shil'nov$^{1,5,}$\thanks{E-mail: shil@kink.univer.kharkov.ua, 
 visit2@ieec.fcr.es} \\ 
[1.5ex]
$^1${\it Consejo Superior de Investigaciones Cient\'{\i}ficas,}\\
{\it IEEC, Edifici Nexus-204, Gran Capit\`a 2-4, 08034, Barcelona, Spain}\\
[0.5ex]
$^2${\it Department  ECM, Faculty of Physics, University of Barcelona,}\\
{\it Diagonal 647, 08028, Barcelona, Spain}\\
[0.5ex]
$^{3}${\it Dept. Fisica, CCET, Universidade Federal de Sergipe,}\\
{\it  490000-000 Aracaju, SE, Brasil}\\
[0.5ex]
$^{4}${\it  Departament of Mathematics and Physics,} \\ 
{\it Tomsk Pedagogical
University, 634041, Tomsk, Russia}\\
[0.5ex]
$^5${\it Department of Theoretical Physics, Faculty of Physics,}\\
{\it Kharkov State University, Svobody Sq. 4, 310077,
Kharkov, Ukraine}
}}
\date{ }
\maketitle
\abstract{A four-fermion model with additional higher-derivative terms
is investigated in an external electromagnetic field.
The effective potential in the leading order of large-$N$ expansion is
calculated in external constant magnetic and electric fields. 
 It is shown that, in contrast to the
former results concerning the universal character of ``magnetic catalysis'' in
dynamical symmetry breaking, in the present higher-derivative  model the
magnetic 
field  restores  chiral symmetry broken initially on the tree level.
Numerical  results describing a second-order phase transition that
accompanies the symmetry restoration at the quantum level are presented.}


\newpage
\section{Introduction}

The dynamical symmetry breaking (DSB) issue has attracted
a lot of attention since the appearence of the seminal paper by  Nambu 
and Jona-Lasinio \cite{NJL}. It is, in fact, 
the most economical way to realize the Higgs mechanism. It  has 
been applied to the study of different models of modern quantum field
theory,\footnote{For details and a list of references see, for example, 
\cite{DSB}.} 
especially  four-fermion ones \cite{NJL, GN}, where it is viewed
as a  
low-energy effective action of strong interactions physics \cite{PhysRep}.

Chiral symmetry (CS) breaking and dynamical fermion mass generation
in four-fermion models have been investigated in the presence of 
 external fields:
electromagnetic \cite{Sch}-\cite{emf}, gravitational 
\cite{gf} - \cite{REV},
 and their combination
\cite{REV}, \cite{emfg}. It has been observed that both a positive 
spacetime curvature and an external 
electric field 
 try to restore chiral symmetry  while a  magnetic field and 
negative  curvature always  break it. Therefore external fields
have been shown to induce some new phase transitions and enrich the model 
phase diagram essentially. 

Higher-derivative  extensions of the NJL model have been discussed
recently owing to different reasons.  First, a model with 
higher-derivative terms in the interaction 
vertex was proposed that exhibits an intersting  
  equivalence with the symplified theory of 
electroweak  interactions \cite{HD1}. 
These terms of the low-energy effective action 
were shown to be essential and even unavoidable
 in the strong coupling regime,
where a nontrivial phase diagram with a policritical point was shown to exist
 \cite{HDAA}.
On the other hand, some regularization schemes based on the introduction of
 additional
higher-derivative terms  into the initial kinetic one were discussed
\cite{HD2}. Furthemore, gravity effects upon the DSB for 
the former variant of high-derivative 
NJL model were investigated as well \cite{HDR}.

In the present paper we study the DSB in the four-fermion model with a 
higher-derivative  kinetic term 
\cite{HD2} in an  external electromagnetic field. It is of
 interest to check up 
if the ``universal rules'' concerning magnetic catalysis of DSB 
and chiral symmetry restoration 
under the influence of electric fields work for this more complicated model.
 
\section{ Effective potential of the higher-derivative model}

We are going to investigate 
a higher-derivative generalization of the NJL-like model in
an external constant electromagnetic field
with the following action:
\begin{equation}
S=\int d^4 x\left\{ \overline{\psi}\left[ \gamma^\mu iD_\mu+
\xi\left( \gamma^\mu iD_\mu\right)^2\right] \psi +
{\lambda \over 2N} \left[ (\overline{\psi}\psi)^2+
(\overline{\psi} i \gamma_5 \psi)^2 \right] \right\}, \label{ac1}
\end{equation}
where the covariant derivative $D_{\mu}$ includes the electromagnetic
potential $A_{\mu}$: 
\begin{equation}
D_{\mu}=\partial_{\mu} -i e A_\mu. \label{D}
\end{equation}
It should be noted that in our  model CS is already broken on the tree level
when $\xi \neq 0$ 
in contrast to the original NJL case.
By introducing the auxiliary fields
\begin{equation}
\sigma=-{\lambda \over N }(\overline{\psi} \psi ), \qquad
\pi=-{\lambda\over N}
\overline{\psi} i \gamma_5 \psi , \label{af}
\end{equation}
we can rewrite the action  as
\begin{equation}
S=\int d^4 x\left\{ \overline{\psi}\left[\gamma^\mu iD_\mu +
\xi\left( \gamma^\mu iD_\mu\right)^2-
\left(\sigma+i\pi\gamma_5\right)\right]\psi 
- {N \over 2\lambda}(\sigma^2+\pi^2)\right\}. \label{ac2}
\end{equation}
Then, the effective action in the large-$N$ expansion is given by
\begin{equation}
{1 \over N } \Gamma_{eff}(\sigma,\pi)=
-\int d^4 x{\sigma^2+\pi^2 \over 2\lambda} -
i\ln \det \left\{i\gamma^\mu (x)D_\mu +
\xi\left( \gamma^\mu iD_\mu\right)^2
-(\sigma+i\gamma_5\pi) \right\}. \label{ea}
\end{equation} 
Here we can put $\pi=0$, because the final expression will depend on
the combination $\sigma^2+ \pi^2$ only. 

Defining the effective potential (EP) as
$V_{eff} = -\Gamma_{eff}/N\!{\displaystyle \int} d^4\! x$, for constant
configurations of $\sigma$ and $\pi$  we get
\begin{equation}
V_{eff}={\sigma^2 \over 2\lambda }+
i \Sp \ln \langle x| [ \gamma^\mu iD_\mu+
\xi\left( \gamma^\mu iD_\mu\right)^2
 -\sigma] |x \rangle . \label{ep1}
\end{equation}
By means of the Green function (GF) which obeys the equation
\begin{equation}
(i \gamma^\mu D_\mu + \xi\left( \gamma^\mu iD_\mu\right)^2
-\sigma)_x 
G(x,x',\sigma)=\delta(x-x'), \label{gf1}
\end{equation}
we obtain the following formula
\begin{equation}
V_{eff}'(\sigma)={ \sigma \over \lambda }-i \Sp G(x, x, \sigma). \label{ep2}
\end{equation}

To find out the GF  $G(x, x', \sigma)$ it is convenient to 
represent the 
higher-derivative operator 
 in the  form
\begin{eqnarray}
 i\gamma^\mu D_\mu+\xi\left( \gamma^\mu iD_\mu\right)^2-\sigma  =
\xi \left( i\gamma^\mu D_\mu - M_1\right) 
\left( i\gamma^\mu D_\mu-M_2\right) ,  \label{hd}
\end{eqnarray}
where
\begin{eqnarray}
M_1 &=&\frac{-1+\sqrt{1+4\xi \sigma}}{2\xi },
\smallskip\ \smallskip\ M_2=\frac{-1-%
\sqrt{1+4\xi \sigma}}{2\xi }.   \label{Mi}
\end{eqnarray}
Here, we suppose that $\xi \geq -1/4\sigma$.
Then we can  represent the GF  as
\begin{equation}
 G(x, x', \sigma)=\left( 1+4\xi \sigma\right) ^{-1/2}
\left\{ S(x, x', M_1)-S(x, x', M_2)\right\}, \label{gf2}
\end{equation}
 where the functions $S(x, x', M_i)$ obey the equation:
\begin{equation}
(i \gamma^\mu D_\mu - M_i)_x S(x, x', M_i)=\delta(x-x') \label{gf3}
\end{equation}
and $i=1,2$.
This is exactly the usual GF of  
massive fermions in a constant 
external electromagnetic field, whose 
proper-time
representation is well-known \cite{Sch}.
We have now derived all  the preliminary formulae 
needed to construct the EP  of our model.

\section{Dynamical symmetry breaking without an external field}

In the absence of an electromagnetic field, the GF (\ref{gf3}) 
in the proper-time 
representation has the form \cite{Sch}:
\begin{eqnarray}
S(x-x', M_i)=
 -\int\limits_{1/\Lambda^2}^\infty
\frac{ds}{(4\pi s)^2}
\exp\left[-is {M_i}^2 -\frac{i}{4s}(x-x')^2\right]
\left[ M_i+\frac{1}{2s}\gamma^\mu (x-x')_\mu \right],
  \label{gf0}
\end{eqnarray}
where $\Lambda$ is an ultraviolet cut-off parameter.
It should be noted that, in contrast to the paper \cite{HD2},
where UV divergences have been dealt with by introducing a
 cut-off as a multiplier of the higher-derivative term, we have to insert 
in our present model this special parameter $\Lambda$ anyway.
The point is that the action of our model contains only squares
 of derivatives and 
this is not enough in order to regularize one-loop Feynman graphs in a
four-dimensional 
spacetime. Meanwhile the kinetic term of the model studied in  \cite{HD2} 
contains a cubic
 higher-derivative term  which provides a complete regularization 
of any diagram.

   The EP generated by the two functions
$S(x, x', M_i)$ is the following
\begin{eqnarray}
V_{eff} (\sigma)=-4i\left( 1+4\xi \sigma\right) ^{-1/2}
\left[ \int\limits_0^{M_1} dm \Sp S(x, x, m)-
\int\limits_0^{M_2} dm \Sp S(x, x, m) \right], \label{v1} 
\end{eqnarray}
where the multiplier 4 is the dimension of the fermion representation,
and the opposite sign of the second integral should be noted.
After a Wick rotation, $is \rightarrow s$,
and integration over $m$,
we get a positive expressions for both  terms  of EP
(\ref{v1}) and  no negative modes appear here.
Therefore the EP can be  written as
\begin{eqnarray}
V_{eff}(\sigma)= \frac{\sigma^2}{2\lambda} +
\frac{1}{8\pi^2}\!\int\limits_{1/\Lambda^2}^\infty
\frac{ds}{s^3}
\left( e^{-sM_1^2}+e^{-sM_2^2}\right).
\end{eqnarray}
Substituting here the expressions (\ref{Mi}) and perfoming the integration 
over $s$, we finally obtain the effective potential (\ref{ep1})
\begin{eqnarray}
V_{eff}(\sigma) &=& \frac{\sigma^2}{2\lambda}+
\frac{1}{16\pi^2 \xi^4}\left\{2\Lambda^4\xi^4-
2 \Lambda^2 \xi^2 \left(1+2\xi\sigma\right)+
(\frac{3}{2}-\gamma)
\left[1+4\xi\sigma+2\left(\xi\sigma\right)^2\right] \right. \nonumber \\
&& \hspace*{-9mm} -\frac{1}{2}\left[1+4\xi\sigma+2\left(\xi\sigma\right)^2-
\left(1+2\xi\sigma\right)\sqrt{1+4\xi\sigma}\right]
\ln\frac{1+2\xi\sigma-\sqrt{1+4\xi\sigma}}{2\xi^2\Lambda^2}
\\   &&
\left. -\frac{1}{2}\left[1+4\xi\sigma+2\left(\xi\sigma\right)^2+  
\left(1+2\xi\sigma\right)\sqrt{1+4\xi\sigma}\right]
\ln\frac{1+2\xi\sigma+\sqrt{1+4\xi\sigma}}{2\xi^2\Lambda^2}\right\},
\nonumber
\end{eqnarray}
where we assume that $\sigma << \Lambda$. 

The gap equation
\begin{equation}
V_{eff}'(\sigma)=0
\end{equation}
can be found directly and CS 
turns out to be broken
for any value of the coupling constant and $\xi >0$,
 because
\begin{equation}
V_{eff}'(0)=\frac{1}{4\pi^2\xi^3}\left[-\Lambda^2\xi^2+1-\gamma+
\ln\left(\Lambda^2\xi^2\right)\right]
\label{v'0.b0}
\end{equation}
is always negative (see Fig. 1).
This is an absolutely natural result because the higher-derivative
term in the action (\ref{ac1})
is noninvariant under chiral transformations, both of continuous and discrete 
type. It causes the presence of a  bare current mass
which is preserved even in the limit $\sigma \to 0$ when $M_2^2 \to 1\!/\xi^2$ 
being  non-zero in all cases.

\section{Symmetry restoration under the influence of external constant 
magnetic or electric fields}

The GF in an external constant magnetic field is given by 
\cite{Sch, mf}:
\begin{eqnarray}
 \hspace*{-20mm} S(x-x', M_i)=
-\int\limits_{1/\Lambda^2}^\infty
\frac{ds}{(4\pi s)^2}
e^{-is {M_i}^2} \exp(-\frac{i}{4s}(x-x')_\mu C^{\mu\nu}(x-x')_\nu)\\
\times\left( M_i+\frac{1}{2s}\gamma^\mu C_{\mu\nu}(x-x')^{\nu}-
\frac{e}{2}\gamma^\mu F_{\mu\nu}(x-x')^\nu\right)
\left( eBs \cot (eBs)-\frac{es}{2}\gamma^\mu\gamma^\nu
F_{\mu\nu}\right),  \nonumber \label{gf.mf}
\end{eqnarray}
where 
\begin{equation}
C_{\mu\nu}=\eta_{\mu\nu} +F_{\mu}{}^{\lambda} F_{\lambda\nu}
{1- eBs \cot (eBs) \over B^2}.
\end{equation}
Then, we can write  the EP 
\begin{eqnarray}
 \hspace*{-20mm} V_{eff}(\sigma)= \frac{\sigma^2}{2\lambda} +
\frac{eB}{8\pi^2}\!\int\limits_{1/\Lambda^2}^\infty
\frac{ds}{s^2}\coth(eBs)
\label{V}\\  \times
\left\{\exp \left[-s\left( \frac{\sqrt{1+4\xi\sigma}+1}{2\xi}\right)^2 \right]+
\exp \left[-s\left( \frac{\sqrt{1+4\xi\sigma}-1}{2\xi}\right)^2 \right]
\right\} 
\nonumber  
\end{eqnarray}
The most reliable method to deal with  the divergences 
 is the cut-off scheme.
We can make the following trick: to write the integral in the EP in the form
\begin{equation}
\int\limits_{1/\Lambda^2}^\infty {ds\over s^3}
e^{-sM_i^2}\left[ (eBs) \coth(eBs) - 1 -\frac{1}{3}(eBs)^2\right] +
\int\limits_{1/\Lambda^2}^\infty {ds\over s^3}e^{-sM_i^2}
\left[ 1 +\frac{1}{3}(eBs)^2\right]
\end{equation}
and to calculate the last one by keeping
$\Lambda$ finite, while the first integral is finite already, 
so that we can  
set $1/\Lambda^2=0$ at the lower limit. Then it appears to be possible
to calculate it like a limit $\mu \to -1$, by using the formula
\begin{equation}
\int\limits_0^\infty dx x^{\mu - 1}e^{-ax}\coth (cx)
= \Gamma (\mu) \left[ 2^{1 - \mu}
(c)^{-\mu}\zeta(\mu , \frac{a}{2c})-a^{-\mu}\right].
\end{equation}
After  integration over $s$, we get
\begin{eqnarray}
V_{eff}(\sigma)&=& \frac{\sigma^2}{2\lambda} + 
\frac{1}{8\pi^2}\left\{\Lambda^4 +\frac{2}{3}\left( eB \right)^2
\left[ \ln \left( \frac{\Lambda^2}{2eB}\right)-\gamma\right]\right.\\
&& -\left.\left( M_1^2+M_2^2 \right) \Lambda^2 + 
\frac{1}{2} \left( M_1^4 + 
M_2^4 \right) \left[ \ln\left(\frac{\Lambda^2}{2eB} \right) +
1-\gamma \right] \right. 
\nonumber \\  && +
eB \left[M_1^2 \ln \left(\frac{M_1^2}{2eB}\right)+
M_2^2\ln \left(\frac{(M_2^2}{2eB} \right) \right]
\nonumber \\ && -
\left. 4(eB)^2\left[\zeta'\left(-1,\frac{M_1^2}{2eB}+1\right)+
\zeta'\left(-1,\frac{M_2^2}{2eB}+1\right)\right]\right\}+
{\cal O}\left(\frac{1}{\Lambda}\right),
\nonumber
\end{eqnarray}
where
\begin{equation}
\zeta'(\nu, x)=\frac{d}{d\!\nu} \zeta(\nu, x).
\end{equation}
Substituting the expressions for $M_1^2$ and $M_2^2$, we obtain the
following formula for the EP
\begin{eqnarray}
V_{eff}(\sigma)&=& \frac{\sigma^2}{2\lambda} +
\frac{1}{8\pi^2}\left\{\Lambda^4 +\frac{2}{3}\left( eB \right)^2
\left[ \ln \left( \frac{\Lambda^2}{2eB}\right)-\gamma\right]\right. \\
&& - \frac{\Lambda^2}{\xi^2}\left(1+2\xi\sigma \right)+
\frac{1+4\xi\sigma+2(\xi\sigma)^2}{2\xi^4}
\left[\ln\left( \frac{\Lambda^2}{2eB}\right)+\frac{3}{2}-
\gamma\right] 
\nonumber \\ && +
\frac{eB}{2\xi^2}\left(1+2\xi\sigma-\sqrt{1+4\xi\sigma}\right)
\ln\frac{1+2\xi\sigma-\sqrt{1+4\xi\sigma}}{4\xi^2eB}  
\nonumber \\  && +
\frac{eB}{2\xi^2}\left(1+2\xi\sigma+\sqrt{1+4\xi\sigma}\right)
\ln\frac{1+2\xi\sigma+\sqrt{1+4\xi\sigma}}{4\xi^2eB} 
\nonumber \\  && -
4(eB)^2\left[\zeta'\!\left(-1, 1+
\frac{1+2\xi\sigma-\sqrt{1+4\xi\sigma}}{2eB\xi^2}\right)\right.
\nonumber  \\  && +
\left.\left.\zeta'\!\left(-1, 1+
\frac{1+2\xi\sigma+\sqrt{1+4\xi\sigma}}{2eB\xi^2}\right)\right]
\right\}
\nonumber
\end{eqnarray}
To see if there are any possibilities for restoration of the chiral
symmetry in this model, one should calculate the derivative
$V_{eff}'(\sigma)$ at the origin $\sigma=0$:
\begin{eqnarray}
V_{eff}'(0)= \frac{1}{8\pi^2\xi^3} \left[ 2\ln ( \Lambda^2\xi^2) -
2 \Lambda^2\xi^2 +3-2\gamma + 2eB\xi^2 - 2 \ln (2eB\xi^2) \right. \\ -
\left.  2eB\xi^2 \ln ( 2eB\xi^2)+ 2eB\xi^2 \ln(2\pi) - 
4eB\xi^2 \ln\!\Gamma\left(1+\frac{1}{ 2eB\xi^2}\right) \right] \nonumber
\label{v'0.b}
\end{eqnarray}
It should be noted that this formula  does not reproduce Eq. 
(\ref{v'0.b0}) in the limit $B \to 0$, due to the circumstance that  
it was actually  calculated as the zero-order term in the power 
expansion on the dimensionless parameter $\sigma^2/eB$. Thus, one has to 
keep $eB$ finite here.

As is clear from Fig.~2, there is a rather big area of values of the 
magnetic field strength
and the cut-off parameter  where the derivative $V_{eff}'(0)$ is
{\it positive}. That indicates the CS restoration on the quantum level. The 
corresponding numerical analysis 
proves in fact that this rerstoration occurs continuously 
with magnetic field strength growth.
This means that there is a second-order phase transition as  shown in Fig.~3.
The same type of phase transition induced by a change of the nonlinearity 
parameter $\xi$  is depicted in Fig.~4.

For the case of an external constant electric field, the EP 
has almost the same form:
\begin{eqnarray}
V_{eff}(\sigma)&=& \frac{\sigma^2}{2\lambda} +
\frac{eE}{8\pi^2}\!\int\limits_{1/\Lambda^2}^\infty
\frac{ds}{s^2}\cot(eEs)
\label{Vef}\\ && \times
\left\{\exp \left[-s\left( \frac{\sqrt{1+4\xi\sigma}+1}{2\xi}\right)^2 \right]+
\exp \left[-s\left( \frac{\sqrt{1+4\xi\sigma}-1}{2\xi}\right)^2 \right]\right\}
\nonumber
\end{eqnarray}
or, after substitution of the expressions for $M_1, \, M_2$,
\begin{eqnarray}
V_{eff}(\sigma) &=& \frac{\sigma^2}{2\lambda} +
\frac{1}{8\pi^2}\left\{\Lambda^4 - \frac{2}{3}\left( eE \right)^2
\left[ \ln \left( \frac{\Lambda^2}{2ieE}\right)-\gamma\right]\right. \\
&& - \frac{\Lambda^2}{\xi^2}\left(1+2\xi\sigma \right)+
\frac{1+4\xi\sigma+2(\xi\sigma)^2}{2\xi^4}
\left[\ln\left( \frac{\Lambda^2}{2ieE}\right)+\frac{3}{2}-
\gamma\right]
\nonumber \\ && +
i\frac{eE}{2\xi^2}\left(1+2\xi\sigma-\sqrt{1+4\xi\sigma}\right)
\ln\frac{1+2\xi\sigma-\sqrt{1+4\xi\sigma}}{4i\xi^2eE}
\nonumber \\ && +
i\frac{eE}{2\xi^2}\left(1+2\xi\sigma+\sqrt{1+4\xi\sigma}\right)
\ln\frac{1+2\xi\sigma+\sqrt{1+4\xi\sigma}}{4i\xi^2eE}
\nonumber \\ && +
4(eE)^2\left[\zeta'\!\left(-1, 1-
i\frac{1+2\xi\sigma-\sqrt{1+4\xi\sigma}}{2eE\xi^2}\right)\right. \\
&& + \left.\left.\zeta'\!\left(-1, 1-
i\frac{1+2\xi\sigma+\sqrt{1+4\xi\sigma}}{2eE\xi^2}\right)\right]
\right\}.
\nonumber
\end{eqnarray}
This expression has an imaginary part defining a particle
creation velocity, and strictly speaking the vacuum becomes unstable  
\cite{GG}.
However, for some small values of the electric field strength, when particle 
creation is still exponentially depressed, we can perform an analysis of
the  DSB
phenomenon using the real part of the EP. 

To estimate if symmetry restoration takes place in an external electric field,
we can use again the value of the derivative of the EP 
 at the origin:
\begin{eqnarray}
\Re V_{eff}'(0)&=& \frac{1}{8\pi^2\xi^3} \left[ 2\ln ( \Lambda^2\xi^2) -
2 \Lambda^2\xi^2 +3-2\gamma \right. \\ && -
\left. 2 \ln (2eE\xi^2) +  \pi eE\xi^2 +  
4eE\xi^2 \arg\Gamma\left(1-\frac{i}{ 2eE\xi^2}\right) \right]. \nonumber
\label{v'0.e}
\end{eqnarray}

Performing now a numerical analysis of the same fashion as in previous section,
we may get qualitatively the same results. There are regions where CS
is restored due to the electrical field effect. Being the procedure very 
similar, we do not present here explicit figures of that analysis for the sake
 of concissness.

\section{Conclusions}

We have studied in this paper the influence of magnetic  and
electric external fields on
the CS restoration in a
 higher-derivative NJL-like model, where this symmetry 
is broken already in 
the absence of external fields. It has been shown that  a domain of
parameter values exists  within the range
of validity of our approximation 
 where the external magnetic or electric fields restore chiral 
symmetry, at least at the quantum level. This is  in 
contrast to the  magnetic catalysis phenomenon occuring
 in the usual NJL model.

Fortunately the phase transition accompanying this symmetry restoration is 
a second order one and its continuous character reasures us on the correctness 
of our approximation. In fact, within the broken phase, for every value of 
$\Lambda$  there is some vicinity of the  origin  where 
$\sigma_{min} << \Lambda$, in accordance with the restrictions under which  we 
have obtained the formula for the effective potential.

 It should also be noted that it is not difficult
to extend our model (and our calculational scheme) to include
other higher derivative terms in the kinetic piece of the
Lagrangian. In particular, one can consider a Lagrangian with a
term of fourth order.

   We would like to thank  A.A. Andrianov for helpful discussions. 
This work has been partly financed by DGICYT (Spain), project PB96-0925, and
by  CIRIT (Generalitat de Catalunya),  grant 1995SGR-00602.
The work of Yu.I.Sh.  was supported in part by Ministerio de
Educaci\'on y Cultura (Spain), grant SB96-AN4620572.
The work by S.D.O. has
been partially supported by MEC (Spain). S.P.G. thanks the Brazilian
foundation CNPq for support.

\pagebreak
\ni{\large \bf Figure captions}
\ms

\ni {\bf Fig. 1:} Plot of the derivative $v\equiv V_{eff}'(0)/\Lambda^3$ of the
effective potential for $\sigma=0$ versus the dimensionless variable $x\equiv 
\Lambda^2 \xi^2$.
The fact that this derivative is always negative proves that chiral symmetry
is always broken as well.
 \ms

\ni {\bf Fig. 2:} Plot of the derivative value  
$V_{eff}'(0)/\Lambda^3$ of the
effective potential as a function of the two variables $x$ and $y\equiv 
eB \xi^2$. When the magnetic field is non-zero and for a reasonable rang
of values of $x$, a domain  is formed where the derivative of the potential is
positive, signaling a phase transition accompanied by chiral symmetry 
restoration.
 \ms

\ni {\bf Fig. 3:} Plot of the effective potential 
$V(s)\equiv [V_{eff} (\sigma) - V_{eff} (0)]/\Lambda^4$ 
as a function of $s\equiv \sigma /\Lambda$, for a particular value of 
$x=.15$ and three different values of $y=.05,.15,.3$, showing that
the phase transition takes place.
 \ms

\ni {\bf Fig. 4:}  Plot of the potential $V(s)$ but now for a 
particular value of $y=.1$ and several values of $x=.1,.2,.3$. 
\pagebreak
\begin{figure}
\psfig{figure=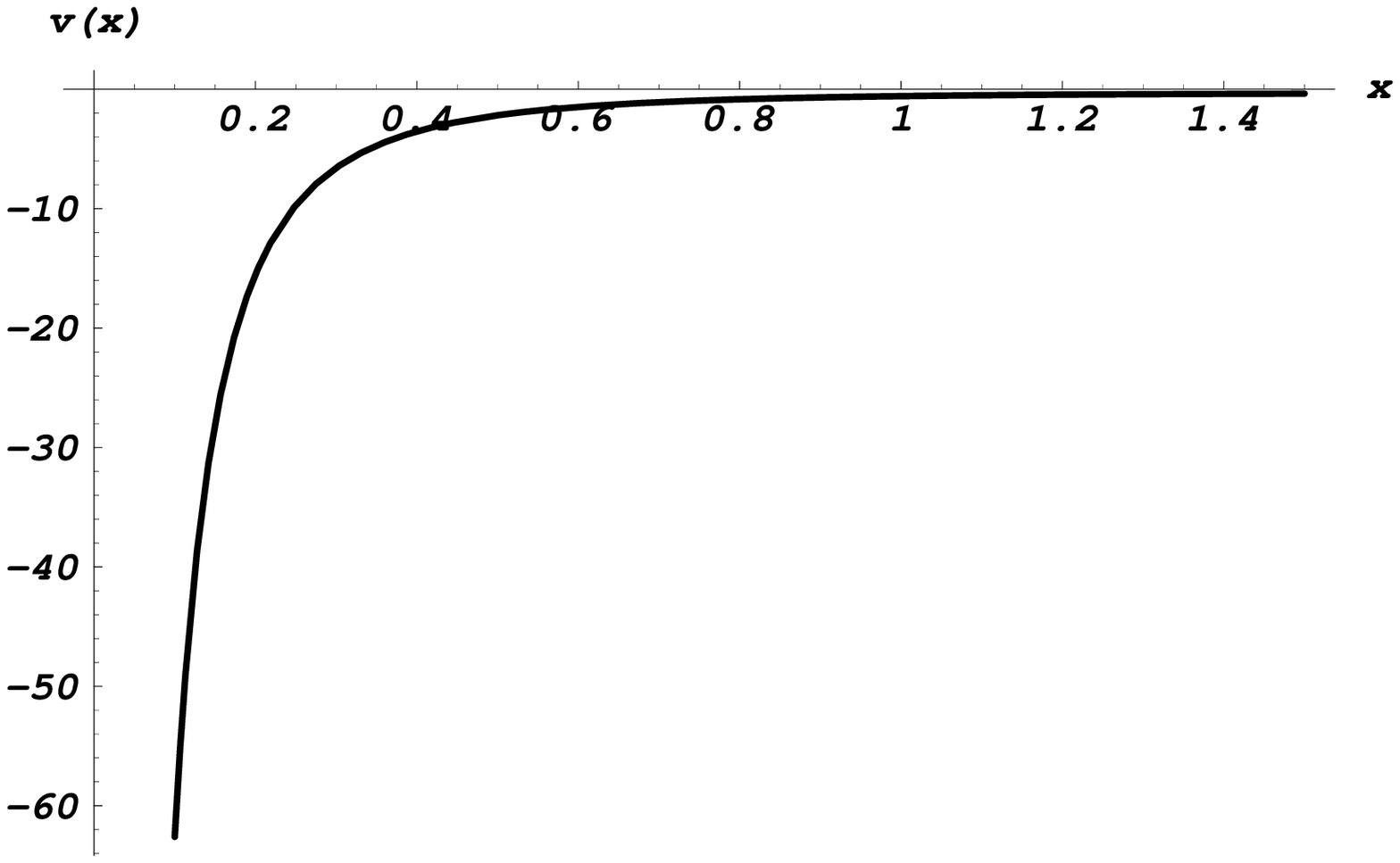}
\begin{center}
{\large Figure 1}
\end{center}
\end{figure}
\pagebreak 
\begin{figure}
\psfig{figure=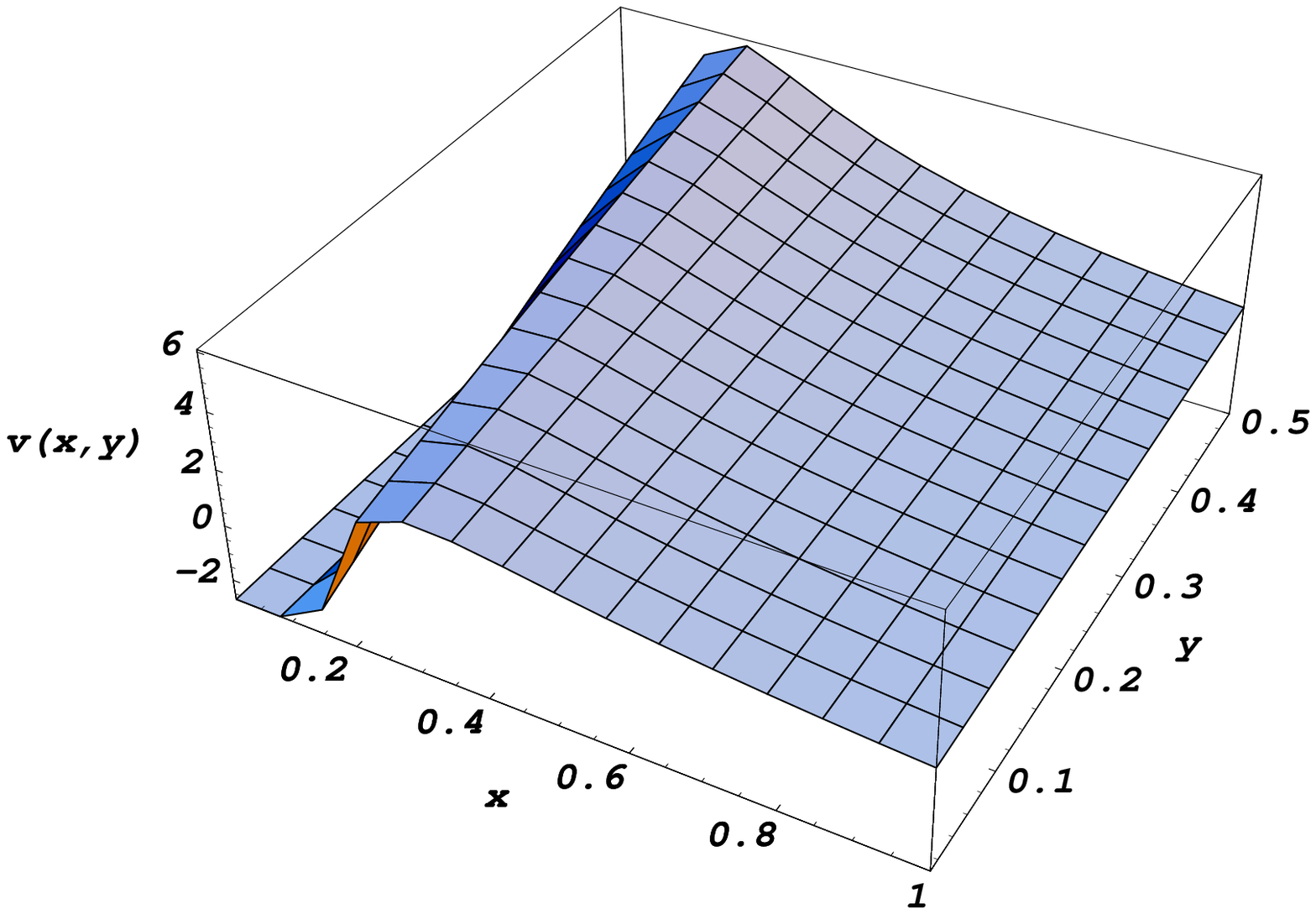} 
\begin{center}
{\large Figure 2}
\end{center}
\end{figure}
\pagebreak
\begin{figure} 
\psfig{figure=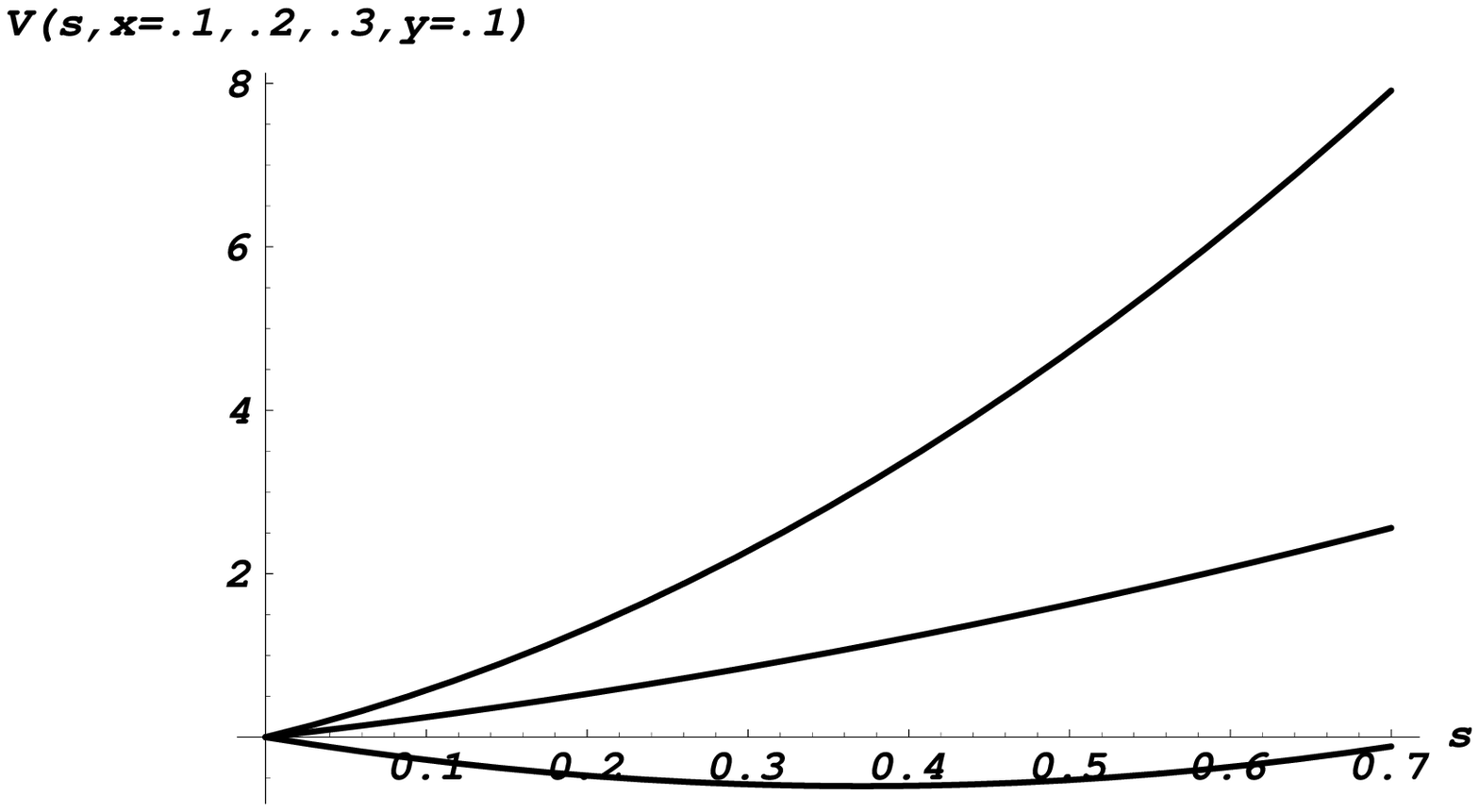}
\begin{center}
{\large Figure 3}
\end{center}
\end{figure}
\pagebreak  
\begin{figure}
\psfig{figure=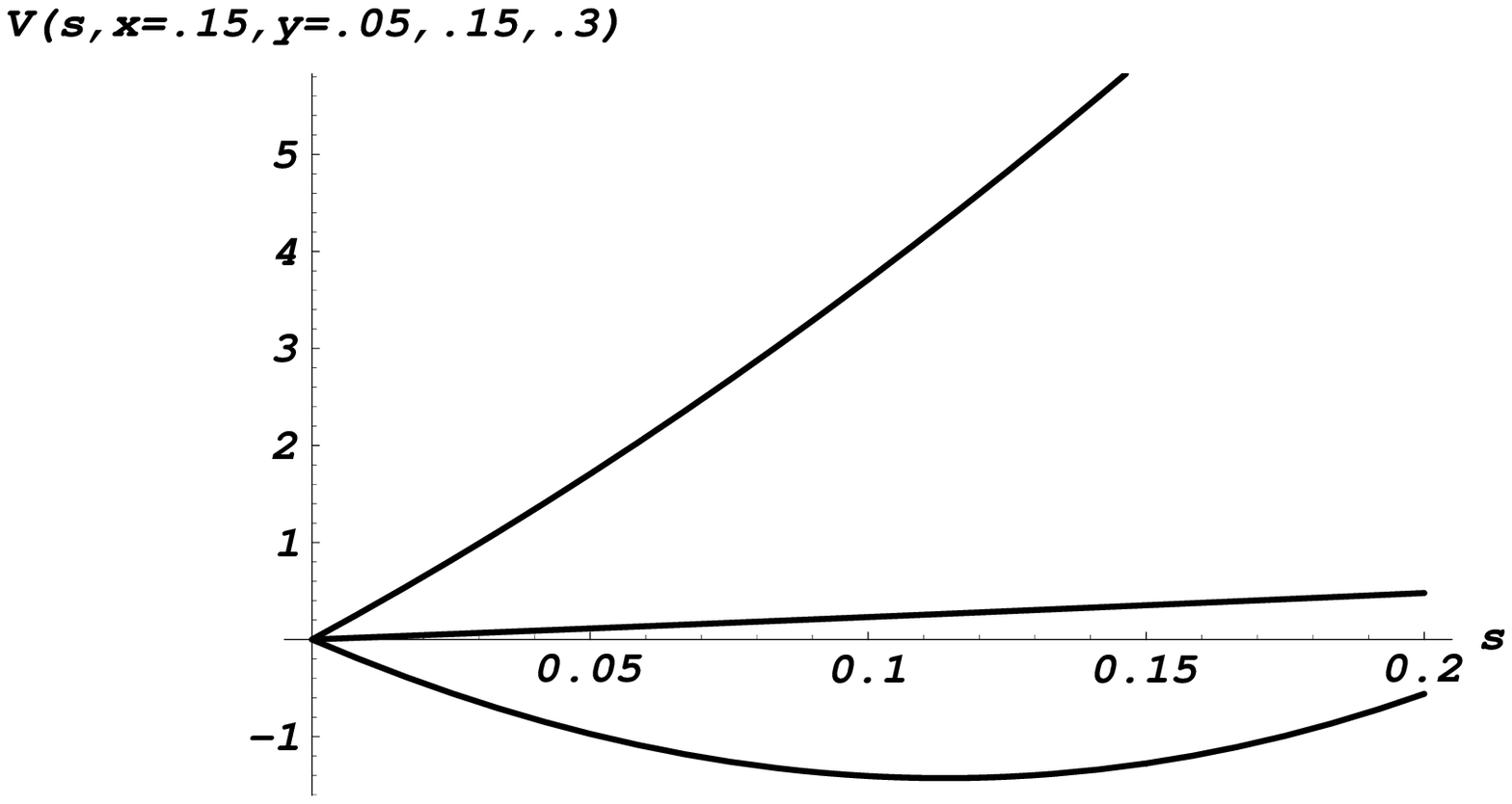}
\begin{center}
{\large Figure 4}
\end{center}
\end{figure}
\end{document}